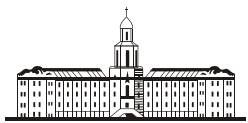

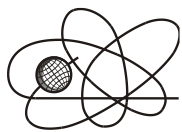

РОССИЙСКАЯ АКАДЕМИЯ НАУК

ИНСТИТУТ ПРОБЛЕМ
БЕЗОПАСНОГО РАЗВИТИЯ
АТОМНОЙ ЭНЕРГЕТИКИ

RUSSIAN ACADEMY OF SCIENCES

NUCLEAR SAFETY
INSTITUTE

Препринт ИБРАЭ № IBRAE-2018-08

Preprint IBRAE-2018-08

**Arutyunyan R.V.**

# THEORETICAL INVESTIGATION OF ELECTRONIC PROPERTIES OF HIGHLY CHARGED FULLERENES. SYSTEMS OF DISCRETE SHORT-LIVED VOLUME-LOCALIZED LEVELS

Москва
2018

Moscow
2018

Abstract

We study the electronic properties of charged fullerenes and onion-like structures in the framework of a simple physical model and show the existence of a system of discrete short-lifetime quantum levels for electrons in the model well potential. In the case of positively charged fullerenes, we find that the energy of the volume-localized levels ranges from 1 eV to 100 eV.

Electrons captured by these discrete levels localized in the volume generate a specific nano-atom wherein electrons are localized inside a charged hollow sphere of fullerene playing the role of a nucleus in an atom.

In case of negatively charged single-layered or onion-like structure fullerenes, Coulomb field creates a spherical potential well for positively charged particles (protons, nuclei of deuterium, tritium or $\mu^+$- mesons). In such a case, a system of discrete levels for positively charged particles is created wherein protons act as electrons and negatively charged sphere of fullerene plays the role of a nucleus.

# Theoretical investigation of electronic properties of highly charged fullerenes. Systems of discrete short-lived volume-localized levels

Rafael V. Arutyunyan
(arut@ibrae.ac.ru)

## 1 Introduction

Fullerenes represent one of allotropes of carbon, along with graphite, diamond, amorphous carbon, nanotubes and graphene. Following the earlier theoretical predictions, the first fullerene $C_{60}$ molecule was experimentally discovered in the 1980-ies [1, 2] as a nanometer-size hollow spherical structure of 60 carbon atoms located at the vertices of a truncated icosahedron. Subsequently, the production of fullerenes in large quantities was developed and the fullerene nanotubes and many other fullerenes were discovered, such as $C_{20}$, $C_{70}$ and even lager structures. This gave a start to an explosive growth of research in the area of nanoscience, the historic development and the current status of which can be found in the numerous reviews [3, 4, 5, 6, 7].

During the recent time, the properties of charged fullerenes have been actively experimentally and theoretically investigated [8, 9, 10, 11, 12, 13, 14, 15, 16, 17, 18, 19, 20, 21, 22]. A considerable number of works are devoted to the study of their stability (lifetime), mechanisms for their charging and decay [23].

The present paper is devoted to the discussion of the structure of the electronic spectrum of the charged fullerenes. Simple models are used to show the existence of the volume-localized discrete quantum levels for the usual fullerene and for the onion-like structures. Here we confine our analysis to the case of the positively charged fullerenes, with a particular attention to the properties of the $C_{60}$ molecule.

Basic notations are as follows: $m_e$ and $e$ are electron's mass and the absolute value of electron charge, $\lambdabar_e = \dfrac{\hbar}{m_e c}$ is the electron Compton length, $\varepsilon_0$ is the electric constant of vacuum; $a_0 = \dfrac{4\pi\varepsilon_0 \hbar^2}{m_e e^2}$ is the Bohr radius, $\alpha = \dfrac{e^2}{4\pi\varepsilon_0 \hbar c}$ is the fine structure constant.

## 2 Charged fullerene: preliminaries

One can multiply ionize $C_{60}$ with the help of the highly charged ions, fast electrons, or photons [7]. An interesting issue is actually how high is the value of an electric charge that a fullerene can carry? Experimentally, charged fullerenes in the range of $Z = 0, \cdots, 9e$, [13], and even up to $Z = 10e$, [11], were produced in collisions of a beam of $C_{60}$ with a beam of highly ionized Xe atoms; such charged fullerenes are stable on a time scale of several $\mu$s. The highest value $Z = 12e$ was observed for a charged fullerene (with the lifetime of of order of a $\mu$s) ionized by intense short infrared laser pulses [12]. The theoretic analysis of the Coulomb stability of highly charged fullerenes [14, 15] predicted the limiting value $Z = 18e$ on the basis of a conducting sphere model, whereas the existence of $Z = 14e$ was established theoretically [16, 17, 18, 19] by means of the density functional theory. However, the predicted lifetime falls drastically --by ten orders-- when $Z$ increases from 11 to 14.

Let us formulate the corresponding quantum-mechanical spectral problem. With an account of the spherical symmetry of a fullerene, we use the standard ansatz for the wave function $\psi(r, \vartheta, \varphi) = R(r) Y_{lm}(\vartheta, \varphi)$, with the spherical harmonics $Y_{lm}$, and recast the spherically symmetric Schrödinger equation [24] into a second order differential equation

$$\frac{d^2 \chi}{dr^2} - \frac{l(l+1)}{r^2} \chi + \frac{2m_e}{\hbar^2}(E - U(r))\chi = 0. \tag{1}$$

Here the function $\chi(r) = rR(r)$, $0 \leq r < \infty$ is subject to the boundary conditions

$$\chi(0) = 0, \quad \chi(\infty) = 0. \tag{2}$$

The form of solution is determined by the potential $U(r)$.

We will discuss the energy levels of an electron by starting from a simple model potential, and then move on to more complicate form of $U$.

## 2.1 Warm-up model: deep spherical well

A first very approximate estimate of the energy levels can be obtained by describing a charged fullerene by the model potential of a spherical rectangular well of the depth $U_0 = \dfrac{Ze}{4\pi\varepsilon_0 R_f}$ (where $Z$ is the value of the charge, and $R_f$ is fullerene's radius):

$$U(r) = \begin{cases} -U_0, & r \leq R_f, \\ 0, & r > R_f. \end{cases} \tag{3}$$

Inside such a well ($0 \leq r \leq R_f$), a non-normalized solution of the Schrödinger equation (1) is described by the spherical Bessel function $\chi = j_l(\xi r / R_f)$ which satisfies the boundary condition at zero $\chi(0) = 0$, whereas outside the well ($R_f < r < \infty$) a solution that satisfies $\chi(\infty) = 0$ is given by the spherical Hankel function $\chi = h_l(i\eta r / R_f)$. The two parameters $\eta$ and $\xi$ are algebraically related,

$$\xi^2 + \eta^2 = \frac{2m_e U_0 R_f^2}{\hbar^2} = 2\alpha \frac{Z}{e} \frac{R_f}{\lambda_e}, \tag{4}$$

and they determine discrete energy levels via

$$E = -\frac{\hbar^2 \eta^2}{2m_e R_f^2} = -U_0 + \frac{\hbar^2 \xi^2}{2m_e R_f^2}. \tag{5}$$

The values of parameters $\eta$ and $\xi$ are fixed by the continuity condition of the wave function at $r = R_f$. For $l = 0$, this yields

$$\eta = -\xi \cot \xi. \tag{6}$$

It is worthwhile to notice that the right-hand side of (4) is essentially greater than 1 for highly charged fullerenes. For example, for $C_{60}^{+10}$ we have $\dfrac{2m_e U_0 R_f^2}{\hbar^2} = 132.5$. Consequently, for highly charged fullerenes $Z \sim 10e$ one can use an approximation of a very deep well $U_0 \gg \dfrac{\hbar^2}{2m_e R_f^2}$ deriving the energy levels from a condition of the vanishing of the wave function at the boundary: $j_l(\xi) = 0$.

## 2.2 Shell model: spherical well with Coulomb tail

The simplest model above obviously provides a very rough approximation. A better understanding is achieved by describing a charged fullerene with the model potential of a sphere with a constant surface charge density:

$$U(r) = -Z\Phi(r), \quad \Phi = \dfrac{e}{4\pi\varepsilon_0} \times \begin{cases} \dfrac{1}{R}, & r \leq R, \\ \dfrac{1}{r}, & r > R. \end{cases} \quad (7)$$

This potential differs from (3) by the charecteristic Coulomb tail outside of the fullerene. As before, here $Z = Ne$ is a positive charge, and $R = R_f$ is the fullerene radius. Our attention will be mainly confined to the $C_{60}$ fullerene, when $R_f = 6.627 a_0$.

The characteristic feature of the corresponding wave functions is that they obviously describe the volume-localized states which are basically confined to the central part of the potential, i.e., to the inner region of the fullerene $r \leq R_f$. The number of such states increases for the potential well becoming deeper, which happens when the charge $Z$ of the fullerene grows.

For a highly charged fullerene with $Z = 10e$, the corresponding spectrum and the wave functions are presented in Fig. 1 and Fig. 2, respectively.

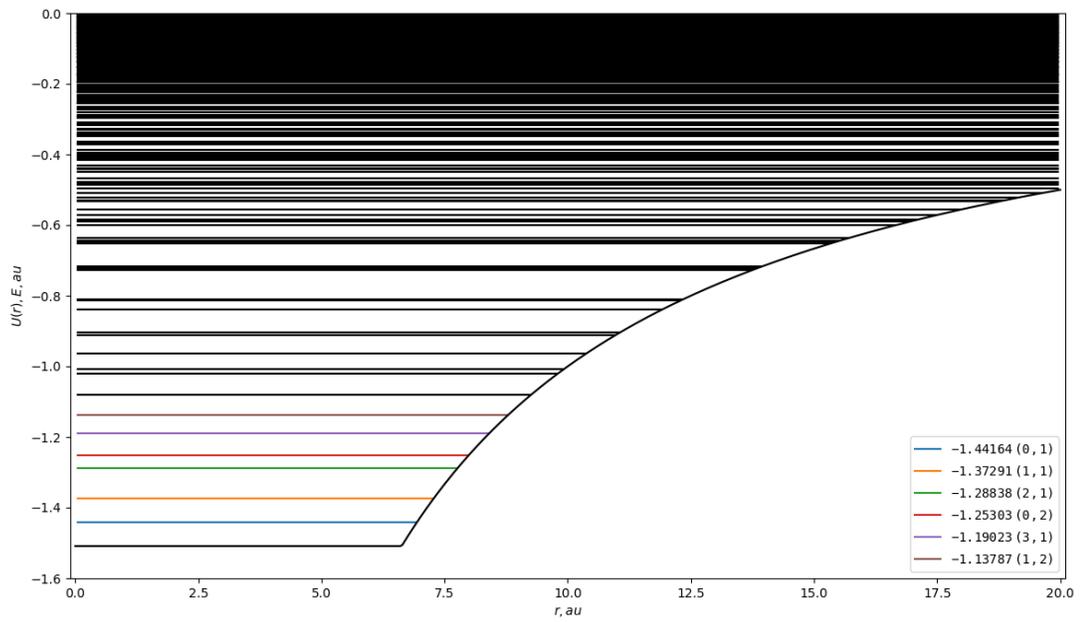

Figure 1: Spectrum for charged fullerene potential (7) with $Z = 10e$.

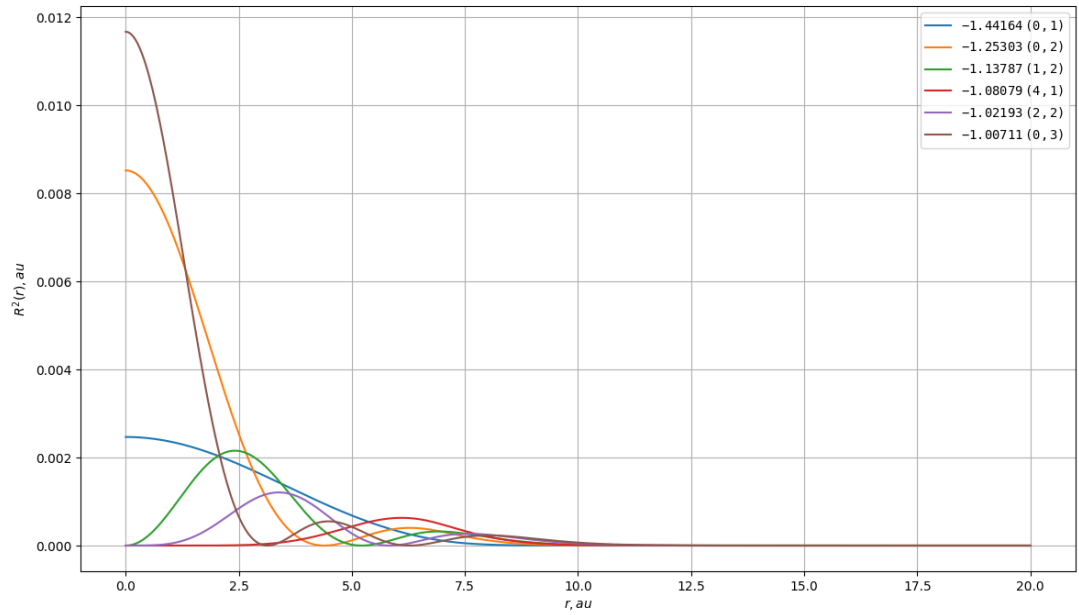

Figure 2: Wave functions for the model potential (7) with $Z = 10e$.

## 3 Model potential for a charged fullerene

The model above provides a rather simplified description in the sense that it does not take into account the actual physical structure of a fullerene. A more realistic potential $U(r)$ can be constructed in the framework of the jellium model [25, 26, 27, 28] as a sum of the positive contribution of the carbon atom's nuclei located on the spherical surface of the fullerene radius $R_f$ and the negative contribution of the electron clouds. The resulting potential is attractive and it has a cusp-shape form with the clear localization in the thin spherical shell. For $C_{60}$, the corresponding Lorentz-bubble potential reads

$$U(r) = -\frac{\frac{\hbar^2}{m_e}V}{(r-R)^2 + d^2}, \qquad (8)$$

where the parameter $V$ determines the depth, $d$ the width, and $R$ the position. In the self-consistent spherical jellium model based on the Kohn-Sham equations, these parameters are fixed [28] to the values

$$V = 0.711, \quad R = 6.627 a_0, \quad d = 0.610 a_0. \qquad (9)$$

In contrast to the volume-localized feature of the wave functions for the model (7), the states for the potential (8) mostly have a typical surface-localized behavior. At the center of a fullerene, the value $U(0) = -0.016$ au $= -0.44$ eV is only slightly below zero, and hence only few discrete levels with the negative energy higher than that value correspond to the volume-localized states.

Coming to the case of a charged fullerene, let us now modify the Lorentz-bubble potential (8) by including the contribution of the charged spherical surface (7). The generalization of the potential (8) for a charged fullerene model then reads

$$U(r) = -\frac{\frac{\hbar^2}{m_e}V}{(r-R)^2 + d^2} - Z\Phi(r), \qquad (10)$$

where $Z$ is the charge of the fullerene.

With such a modification, the central part of the potential deepens, so that $U(0) = -1.52$ au $= -41.5$ eV for $Z = 10e$. As a result, there are two types of wave functions for the modified potential (10): the lower-energy states are distinctly surface-localized, whereas the higher energy levels correspond to the volume-localized quantum states.

One can find the discrete quantum energy levels of an electron in the potential (10) by integrating the Schrödinger equation (1) numerically. The corresponding results for the charge $Z = 10e$ are presented in Table 1 and Figs. 3-5.

Table 1: Electron energy levels for charged fullerene potential (10) with $Z=10e$.
[Notation: $n$ - level number, $l$ - angular quantum number, $i$ - radial quantum number].

| $n$ | $l$ | $i$ | $E$ (au) | | | $n$ | $l$ | $i$ | $E$ (au) |
|---|---|---|---|---|---|---|---|---|---|
| 1 | 0 | 1 | −2.39903 | | | 7 | 6 | 1 | −1.90343 |
| 2 | 1 | 1 | −2.37490 | | | 8 | 7 | 1 | −1.74240 |
| 3 | 2 | 1 | −2.32685 | | | 9 | 8 | 1 | −1.56053 |
| 4 | 3 | 1 | −2.25520 | | | 10 | 0 | 2 | −1.49334 |
| 5 | 4 | 1 | −2.16040 | | | 11 | 1 | 2 | −1.41574 |
| 6 | 5 | 1 | −2.04296 | | | 12 | 9 | 1 | −1.35854 |

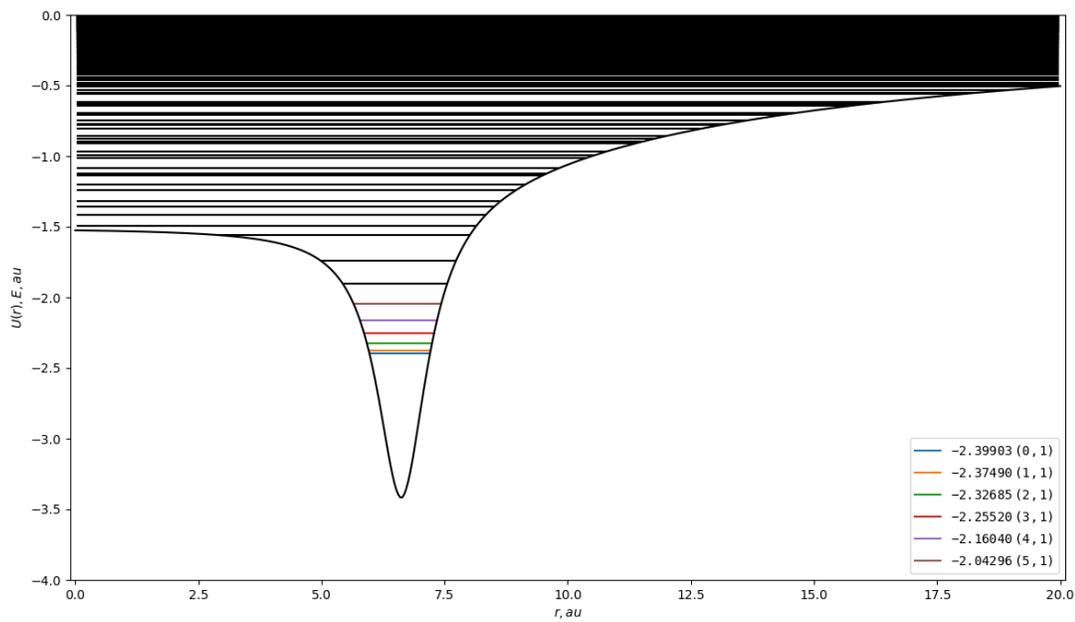

Figure 3: Spectrum for charged fullerene potential (10) with $Z=10e$.

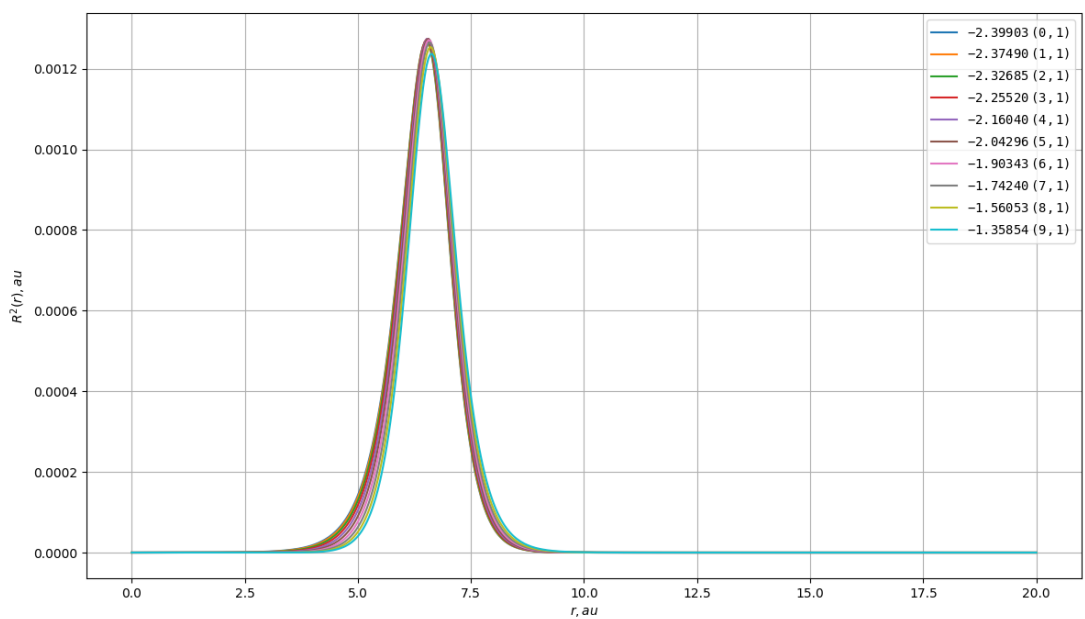

Figure 4: Surface-localized wave functions for charged fullerene potential (10) with $Z = 10e$.

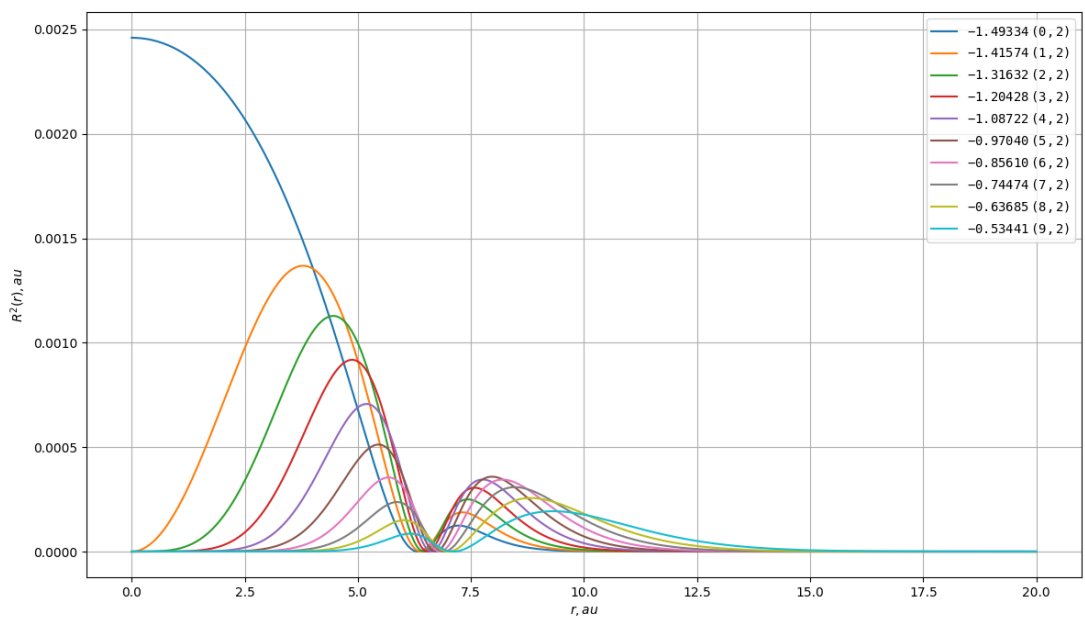

Figure 5: Volume-localized wave functions for charged fullerene potential (10) with $Z = 10e$.

## 4 Onion-like fullerene

Onion (or onion-like) structures represent a highly interesting class of carbon systems which are obtained when fullerenes are concentrically enclosed one into another to form a double-, triple-, or in general a multi-layered object. Each (quasi-)spherical layer is a fullerene ($C_{60}$, $C_{240}$, $C_{540}$, $C_{960}$, $\cdots$), with the separation between shells equal to $0.335$ nm which is slightly larger than the distance between the planar layers in graphite crystals. The outer diameter of a typical 5-15 layered onion ranges between 4 to 10 nm, with the radius of the innermost layer equal to $R_f$, however smaller and much larger onion-like structures are also observed. Following the first observation [29] of spherical multi-layered structures, the concept of ``carbon onion'' was coined in 1992 when the formation of the onion-like spherical particles was demonstrated by heating of nanotubes with an electron beam [30, 31, 32]. Since then the physical and chemical characteristics of carbon onions was analyzed in numerous theoretical and experimental studies; see the reviews [33, 34, 35, 36, 37, 38, 39, 40, 41] for the further information on the production, geometrical, physical and chemical properties, and applications of onion-like carbon structures.

In the context of the current investigation of the energy spectrum of charged carbon complexes, the onion-like structures are qualitatively different from the usual fullerenes in the sense that, in contrast to the latter case when the electric charge is smeared only over the surface of fullerene's hollow sphere, in the former case the electric charge is distributed in the volume of an onion sphere on its many internal layers.

Accordingly, in a simplest model for the study of discrete volume levels of electrons in multi-layer onion-like charged fullerenes (somewhat similarly to the simplest model (5) and (6) of a rectangular spherical well for a fullerene), one can look for analytic estimates by assuming a homogeneous density when the charges of consecutive layers of the onion structure are proportional to the cube of the layer radius. In this case, the potential energy of an electron in an electrostatic field is as follows:

$$U(r) = -\frac{Ze}{4\pi\varepsilon_0} \times \begin{cases} \frac{1}{2R}\left(3 - \frac{r^2}{R^2}\right), & r \leq R, \\ \frac{1}{r}, & r > R. \end{cases} \quad (11)$$

Here $Z$ is the total positive charge of an onion structure, and $R = R_{on}$ is its outer radius. As a first step to understand the spectrum structure, we approximate the potential by extending the piece inside the sphere $r \leq R$ to all values of the radius:

$$U(r) = -U_0 + \frac{1}{2}m_e\omega^2 r^2, \quad (12)$$

where we denoted

$$U_0 = \frac{3}{2}\frac{Ze}{4\pi\varepsilon_0 R_{on}}, \quad \omega = \sqrt{\frac{Ze}{4\pi\varepsilon_0 m_e R_{on}^3}}. \quad (13)$$

The maximal specific charge $Z/N_{tot}$ of onion-like structures (where $N_{tot}$ is the total number of atoms) before their decay would be smaller than that for $C_{60}$, but the absolute value of the charge can be much larger. Accordingly, the depth of the potential well $U_0$ of an electron in the field of a positively charged onion structure then can reach the values of order of 100 eV, thereby increasing the significance of the volume-localized quantum states.

For the approximate potential (11), one can evaluate the energy spectrum analytically by making use of the well-known solution of the Schrödinger equation for the spherical oscillator [24]. For energy levels we find

$$E = E_0 + \hbar\omega(2i + l - 2), \tag{14}$$

$$E_0 = \frac{3}{2}\sqrt{\frac{Ze}{4\pi\varepsilon_0 R_{on}}}\left(\sqrt{\frac{\hbar^2}{m_e R_{on}^2}} - \sqrt{\frac{Ze}{4\pi\varepsilon_0 R_{on}}}\right), \tag{15}$$

whereas the wave functions of the corresponding stationary states are

$$\psi_{nlm} = \text{const} \, r^l \exp\left(-\frac{\lambda r^2}{2}\right) Y_{lm}(\theta, \varphi) \,_1F_1(1 - i, l + \frac{3}{2}, \lambda r^2), \tag{16}$$

where $_1F_1$ is the degenerate hypergeometric function,

$$\lambda = \frac{m_e \omega}{\hbar} = \sqrt{\frac{Zem_e}{4\pi\varepsilon_0 \hbar^2 R_{on}^3}}, \tag{17}$$

the radial quantum number $i = 1, 2, \cdots$, the angular quantum number $l = 0, 1, 2, \cdots$, and $m = 0, \pm 1, \cdots, \pm l$.

As a particular application, let us consider a model of a 5-layer charged onion fullerene with $Z = 225e$ and the size $R_{on} = 5R_f$. The total charge arises from the assumption of a homogeneous distribution of the electric charge on the inner layers proportionally to the third power radius of the layer. The corresponding energy levels (14) for such onion model are presented in Table 2. The well-known degeneracy properties of an oscillator spectrum are manifest.

Table 2: Electron energy levels (14) for an onion-like fullerene potential (11) with $Z = 225$ and $R_{on} = 33.135 a_0$. [Notation: $n$ - level number, $l$ - angular quantum number, $i$ - radial quantum number].

| $n$ | $l$ | $i$ | $E$ (au) | | | $n$ | $l$ | $i$ | $E$ (au) |
|---|---|---|---|---|---|---|---|---|---|
| 1 | 0 | 1 | −10.0676 | | | 11 | 3 | 2 | −9.67442 |
| 2 | 1 | 1 | −9.98900 | | | 12 | 1 | 3 | −9.67442 |
| 3 | 2 | 1 | −9.91035 | | | 13 | 6 | 1 | −9.59578 |
| 4 | 0 | 2 | −9.91035 | | | 14 | 4 | 2 | −9.59578 |
| 5 | 3 | 1 | −9.83171 | | | 15 | 2 | 3 | −9.59578 |
| 6 | 1 | 2 | −9.83171 | | | 16 | 0 | 4 | −9.59578 |
| 7 | 4 | 1 | −9.75307 | | | 17 | 7 | 1 | −9.51714 |
| 8 | 2 | 2 | −9.75307 | | | 18 | 5 | 2 | −9.51714 |
| 9 | 0 | 3 | −9.75307 | | | 19 | 3 | 3 | −9.51714 |
| 10 | 5 | 1 | −9.67442 | | | 20 | 1 | 4 | −9.51714 |

To complete the discussion of the onion-like structures, it is important to analyze the possible values of $Z$ for onion fullerenes. The corresponding experimental and theoretical results on the limiting values of the positive electric charge for ionized onion-like fullerenes are absent. Nevertheless, one can make some simple estimates for the highest value of the charge by evaluating the critical value of the field strength on the outer spherical layer of an onion fullerene. In order to do this, we start with the case of an ordinary fullerene and notice that, for a positive charge $Z$ on it, the value of the electric field strength reads $\mathcal{E} = \frac{Z}{4\pi\varepsilon_0 R_f^2}$, evaluated at the spherical surface of a fullerene, before it becomes unstable due to the field ion emission. Taking $R_f = 6.627 a_0$ and $Z = 12e$ in the fullerene $C_{60}$, we find for the field strength $\mathcal{E} = 1.38 \times 10^{11}$ V/m. This is smaller than the critical value (evaporation field) for the carbon $\mathcal{E}_{max} = 1.48 \times 10^{11}$ V/m, avove which the ion field emission starts [42, 43]. One can reasonably assume that this field value should not be exceeded also for the onion-like structures.

We thus can formulate a simple criterion $\mathcal{E} \sim \mathcal{E}_{max}$ for the stability of a charged multi-layer onion fullerene, with the help of which one can derive a rough estimate of the corresponding maximal possible total positive charge $Z = Ne$. In particular, applying this scheme to the 5-layer onion structure with $R_{on} = 5R_f$ and $N = \sum_i N_i = 225$, we find $\mathcal{E} = 1.08 \times 10^{11}$ V/m which is well below the threshold value. In a similar way, one can evaluate the limiting charge for onion-like structures with an arbitrary number of layers.

Obviously, such a semi-empirical estimate is very approximate and needs to be further refined on the basis of the microscopic calculations or the experimental measurements.

# 5 Conclusions

In the framework of a simple physical model, we demonstrate the existence of a system of discrete short-lifetime quantum levels for electrons in the potential well of the self-consistent Coulomb field of charged fullerenes and onion-like structures. For electrons, in the case of positively charged fullerenes and onion-like structures, the energy of the volume-localized levels ranges from 1 eV to 100 eV.

The results obtained provide a consistent qualitative picture both for the charged fullerenes and for the onion-like structures. In order to refine our very approximate findings, one certainly needs a further investigation on the basis of the microscopic calculations, as well as the experimental measurements.

An experimental confirmation of the existence of the volume-localized discrete levels would be of great interest for the experimental research and practical problems including a development of the new sources of coherent radiation in a wide range of wavelengths.

To estimate the inverted population density $n_{ij}$ necessary to achieve the coherent emission generation threshold on volume-localized levels of fullerenes, we use the following simple expression:

$$\mu_\omega L_{погл} \gg 1 \qquad (1)$$

where:

$$\mu_\omega = \frac{\lambda^2}{2\pi} n_{ij} \frac{\Delta\omega}{\Delta\omega_{sp}} \qquad (2)$$

$\mu_\omega$ is the resonance amplification factor per unit length;
$L_{abs}$ is the photon loss length;
Δω is full broadening of the emission line due to the Doppler effect, collisional broadening and broadening due to nonradiative losses;
$\Delta\omega_{sp}$ is the width of dipole spontaneous emission line; and
ω is the emission frequency at the transition i-j.

$$n_{ij} \gg \frac{2\pi}{\lambda^2 L_{abs}} \frac{\Delta\omega_{sp}}{\Delta\omega} \qquad (3)$$

$$\Delta\omega \approx \Delta\omega_{dop} + \Delta\omega_{col} + \Delta\omega_{без}$$

The value $L_{abs} \sim \frac{1}{n\sigma_{ab}}$, where n is the volume density of fullerenes. According to experimentally measured values in fullerene pairs of $C_{60}$ [ 44 ]

$$\sigma_{abs} \sim 10^{-15} \text{ cm}^2 \text{ in the wavelength range of } 200 - 400 \text{ nm}.$$

An estimate for the volume density of the inverted population $n_{ij}$ at the fullerene density n=$10^{14}$: $10^{17}$ cm$^{-3}$ gives the value of $10^{12}$ - $10^{14}$ cm$^{-3}$

I thank P. N. Vabishchevich for performing numeric computations, and P. S. Kondratenko, Yu. N. Obukhov and participants of the seminar of the Theoretical Physics Laboratory, Institute for Nuclear Safety (IBRAE) for the fruitful and stimulating discussions.